\documentclass[aps,pre,twocolumn,superscriptaddress,floatfix]{revtex4}
\usepackage{graphicx}
\usepackage{amsmath,amssymb}
\usepackage[utf8]{inputenc}
\newcommand{\tkt}{T_{\scriptstyle\rm KT}}

\newcommand{\dmult}{\Delta_{\scriptstyle\rm mult}}

\begin{document}  
\title{Kosterlitz-Thouless and Potts transitions in a generalized XY model}

\author{Gabriel A. Canova}
\affiliation{Instituto de F\'\i sica, Universidade Federal do 
Rio Grande do Sul, CP 15051, 91501-970 Porto Alegre RS, Brazil} 
\author{Yan Levin}
\affiliation{Instituto de F\'\i sica, Universidade Federal do 
Rio Grande do Sul, CP 15051, 91501-970 Porto Alegre RS, Brazil} 
\author{Jeferson J. Arenzon}
\affiliation{Instituto de F\'\i sica, Universidade Federal do 
Rio Grande do Sul, CP 15051, 91501-970 Porto Alegre RS, Brazil}

\date{\today}

\begin{abstract}
We present extensive numerical simulations of a generalized XY model
with nematic-like terms recently proposed by Poderoso {\it et al} [PRL 106(2011)067202].
Using finite size scaling and focusing on the $q=3$ case, we locate 
the transitions between the paramagnetic (P), the nematic-like (N) and the ferromagnetic
(F) phases.  The results are compared
with the recently derived lower bounds for the P-N and P-F transitions.
While the P-N transition is found to be very close to the lower bound, the P-F transition occurs significantly
above the bound.  Finally, the transition between the nematic-like and the ferromagnetic
phases is found to belong to the
3-states Potts universality class. 
\end{abstract}

\maketitle

\section{Introduction}

Two dimensional systems with short-range interactions can not break continuous symmetry. 
This is the reason why neither true crystals, ferromagnets, or nematics can exist in 2D~\cite{MeWa66}.   
Nevertheless, at low temperatures these systems can 
exhibit a quasi-long range order, with the correlation functions
decaying with distance as a power law.  
The transition between the ``ordered'' and disordered phases is often found to be of
infinite order and to belong to the 
Kosterlitz-Thouless (KT) universality class~\cite{KoTh73,Kosterlitz74}. The transition is driven by the unbinding of 
topological charges (vortices).  At low temperature the charges are bound in dipolar vortex-antivortex
pairs, while at high  
temperature the vortices unbind and lead to the destruction of the quasi long-range order.
Unlike the usual thermodynamic phases, the low temperature KT phase 
is critical for all temperatures and is characterized
by a discontinuous jump of the helicity modulus, which is the order parameter that measures
how the system responds to a global twist~\cite{FiBaJa73,NeKo77,OhJa79}. 

A generalization of the XY model, including nematic-like terms, has been recently introduced 
and studied by several authors for both $q=2$~\cite{Korshunov85,LeGr85,LeGrTo86,Hinbergen86,SlZi88,CaCh89,ZuIdTa01,MeNa06,PaOnNaHa08,KoHl10,DiHl11,ShLaFe11,HuWe13} 
and integers $q>2$~\cite{Romano06,PoArLe11},
\begin{equation}
{\cal H} = - \sum_{\langle i j \rangle} [\Delta\cos(\theta_i - \theta_j) +  (1 - \Delta) \cos(q\theta_i - q\theta_j)],
\label{eq.H}
\end{equation}
with $0\leq\Delta\leq 1$, $0\leq\theta_i\leq 2\pi$ and nearest neighbors interactions. 
Similar models have recently been considered in the contexts of collective
motion of active nematics~\cite{NgGiCh12} and Hamiltonian mean field models~\cite{TeBePaLe12}. 
For $\Delta=1$, one recovers the usual XY model with the critical temperature
$\tkt(1)\simeq 0.893$. Changing variables in the partition function, $q \theta_i \rightarrow \bar \theta_i$, 
shows that the $\Delta=0$ model is also isomorphic to the XY model with the same
critical temperature (see also Ref.~\cite{Carmesin87}), $\tkt(0)=\tkt(1)$. In between,
when $0<\Delta<1$, Eq.(\ref{eq.H}) describes the
competition between directional and nematic-like alignment (i.e., $2k\pi/q$ with integer $k\leq q$), with a line
of critical points $\tkt(\Delta)$.
In general, besides the high temperature, paramagnetic (P) phase, there are at least two
other phases that are extensions of the phases occurring at $\Delta=0$
and $\Delta=1$. A quasi long-range ferromagnetic (F) phase exists for all values of $\Delta$ and
extends down to zero temperature (because parallel spins minimize both terms of the Hamiltonian, while
nematic-like ordering minimizes only the second term). For small values of $\Delta$, there is an intermediate 
temperature phase with nematic-like (N) quasi long-range order. A second order phase
transition line is found to separate F and N phases, 
ending in a multicritical point at $\dmult$. Only recently the phase diagram for $q=2$ has 
been precisely obtained~\cite{HuWe13}.  In a recent paper, Poderoso {\it et al} have 
studied~\cite{PoArLe11}
the $q$-nematic N to F transition for the $q=3$ model. This transition was found to belong to the 3-states
Potts universality class. The P-N and P-F 
transitions were expected to belong to the KT class, however, neither the  
location nor the universality class of these transitions was precisely determined in Ref.~\cite{PoArLe11}.
Using heuristic arguments Korshunov~\cite{Korshunov12}, suggested that for $q>2$  a $q$-nematic phase is 
impossible~\footnote{In view of our unpublished Reply (Canova {\it et al}, 2012 arXiv/1207.3447), which is the basis of the present paper, 
Prof. Korshunov withdrew his Comment which, however, is still available on arXiv.}.  
This, however,  clearly contradicted the results of the simulations of Poderoso {\it et al}~\cite{PoArLe11}.  
Furthermore, the mapping between $\Delta=0$ and 1 shows that the N phase exists at least for $\Delta=0$.
Using Ginibre's inequality~\cite{Ginibre70} it can then be shown~\cite{Romano06} that the N-P transition   
must also extend to finite $\Delta$. Ginibre's inequality also provides a rigorous
lower bound~\cite{Romano06} for the transition temperature between P and N phases,  
$\tkt(\Delta) \ge (1-\Delta) \tkt(0)$ for $\Delta<\dmult$.  For $\Delta>\dmult$, the KT transition
is between  P and F phases~\cite{Romano06}, with the lower bound given by 
$\tkt(\Delta)\geq \tkt(0)\Delta$.
Since at very low temperature
the system must be in the F phase, this proves the existence of all three 
phases for small, but finite $\Delta$. The objective of the present work
is to precisely calculate the phase diagram for the $q=3$ generalized XY model and to compare
the critical temperatures for the P-N and P-F transitions with the  
bounds obtained in Ref.~\cite{Romano06}.
Moreover, since the very existence of the N-P transition has been contested for $q>2$~\cite{Korshunov12}, 
it is important to
present a broad set of solid evidences supporting such a transition.


\section{Simulations}
\label{section.simulation}

The simulations were performed on a square lattice of linear size $L$ and
periodic boundary conditions. Both Metropolis single-flip and Wolff cluster algorithms~\cite{Wolff89}
were used. The phase transitions are characterized by observables such as the 
generalized magnetizations and
the corresponding susceptibilities, 
\begin{align}
m_k    &= \frac{1}{L^2}\left| \sum_i \exp(ik\theta_i)\right| \label{eq.magnetization} \\
\chi_k &= \beta L^2 (\langle m_k^2 \rangle-\langle m_k\rangle^2) \label{eq.susc}
\end{align}
where $k = 1,\ldots, q$, and the Binder cumulants~\cite{Loison99,Hasenbusch08},
\begin{equation}
U_k =  \frac{\langle m_k^2 \rangle^2}{\langle m_k^4 \rangle}.
\label{eq.binder}
\end{equation}
Since there is no long-range order in 2d, the observables $m_k$ are not, strictly speaking, the order
parameters for the phase transition.   The order parameter for KT transition is the 
helicity modulus $\Upsilon$ which, in the thermodynamic limit,  is zero in the disordered phase and remains finite 
in the ordered phase. It is defined as the response upon a small, global twist along one particular direction.
Following Ref.~\cite{MiKi03}, it can be written as $\Upsilon =  e -L^2\beta s^2$, 
where $e\equiv L^{-2} \sum_{\langle ij\rangle_x} U_{ij}''(\phi)$ and  $s\equiv L^{-2} \sum_{\langle ij\rangle_x} U_{ij}'(\phi)$
(the sum is over the nearest neighbors along the horizontal direction), $\phi=\theta_i-\theta_j$ and $U_{ij}(\phi)$
is the potential between spins $i$ and $j$. For the Hamiltonian Eq.~(\ref{eq.H})~\cite{HuWe13},
\begin{align}
\Upsilon &= \frac{1}{L^2} \sum_{\langle ij\rangle_x} \left[ \Delta\cos\phi+q^2(1-\Delta)\cos(q\phi)\right] \nonumber \\
         &- \frac{\beta}{L^2}  \left(\sum_{\langle ij\rangle_x}\left[\Delta\sin\phi+q(1-\Delta)\sin(q\phi)\right]\right)^2.
\label{eq.helicity}
\end{align}
For the XY model with $\Delta=1$, the critical temperature is determined by the condition  
$\Upsilon(\tkt)=2\tkt/\pi$ \cite{FiBaJa73,NeKo77}. The isomorphism between the ferromagnetic XY model with $\Delta=1$
and a purely $q$-nematic model with $\Delta=0$ 
requires that the critical temperature must be the same for both models.
The helicity modulus for the $q$-nematic, however, contains an extra factor of $q^2$ which must be accounted for
in the condition for criticality.  Following Ref. \cite{HuWe13}
the location of the KT transition
will be determined by the asymptotic crossing point of $\Upsilon$ and the line $2\tkt/\lambda^2\pi$. 
The factor $\lambda$
is related to the charge of the topological excitation. For the F-P transition ($q=1$), $\lambda=1$ and the
transition is driven by the unbinding of integer vortices. For $q=2$, $\lambda=1/2$ in the N phase, corresponding to half-integer 
vortices (connected by domain walls). In general, $\lambda=1/q$ for the $q$-nematic-paramagnetic transition. 

For the N-F transition
the usual finite size scaling (FSS) analysis provides the critical exponents $\beta, \gamma,$ and $\nu$:
$m = L^{-\beta/\nu} f(t L^{1/\nu})$  and $\chi = L^{\gamma/\nu} g(t L^{1/\nu})$, where $m$ is the
order parameter and $\chi$ is its susceptibility,  $f$ and $g$ are the scaling
functions,  and $t=T/T_c-1$ is the reduced temperature. Indeed, for $q=3$ this transition is in the universality
class of the 3-states Potts model with $\nu=5/6$, $\beta=1/9$ and $\gamma=13/9$, and has been studied
in detail in Ref.~\cite{PoArLe11}. However, the KT transition has an essential singularity, the
correlation length grows exponentially, and
this FSS is no longer valid. Furthermore, the whole low temperature phase is critical and both the correlation length and
the susceptibility are infinite in the thermodynamic limit~\cite{Kosterlitz74,ToCh79}. 
Nevertheless, for all temperatures $T\leq\tkt$,
the critical exponent {\it ratios} are well defined and the magnetization and the associated  susceptibility scale, 
respectively, as $m \propto L^{-\beta/\nu}$ and $\chi \propto L^{\gamma/\nu}$.
Exactly at the transition, $\beta/\nu=1/8$ and $\gamma/\nu=7/4$, which are the same ratios as for the 2d Ising model.
Below the phase transition temperature the critical exponents are non-universal. 

\begin{figure}[hbt]
\includegraphics[width=8cm,height=6cm]{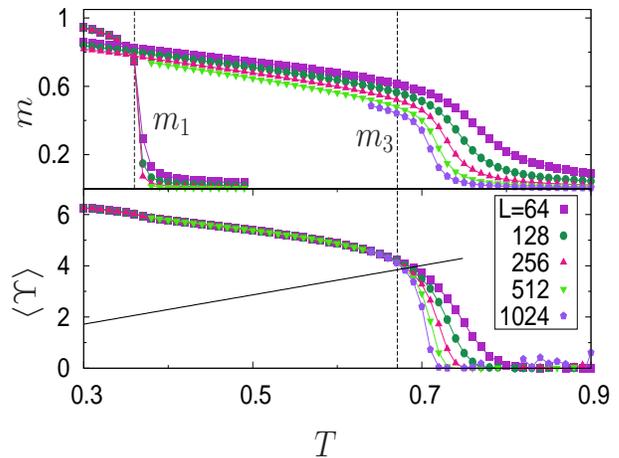}
\caption{(Top) Average magnetization $m_1$ and $m_3$~\cite{PoArLe11} for $\Delta=0.25$ and
several system sizes $L$ showing the N-F phase transitions
at $T_{\rm Potts} \simeq 0.36$ and P-N transition at $T_{\scriptstyle\rm KT}\simeq 0.67$. 
(Bottom) Average helicity modulus $\Upsilon$ versus $T$. The crossing of the helicity with the line $18T/\pi$ (see
text for an explanation) at $T_{\scriptstyle\rm KT}(L)$
gives, for $L\to\infty$,  $T_{\scriptstyle\rm KT}\simeq 0.67$. Notice also that the system sizes used
here are considerably larger than those in Ref.~\cite{PoArLe11}. 
}
\label{fig.helicity}
\end{figure}

Results for the average magnetizations $m_1$ and $m_3$ are shown in Fig.~\ref{fig.helicity} (top)
for $\Delta=1/4$. For this $\Delta$ the critical temperature for the N-F transition was found to be
$T_{\rm Potts}(0.25) \simeq 0.365$~\cite{PoArLe11} at which $m_1$ drops to very low values.
On the other hand, the nematic magnetization $m_3$ clearly shows the N-P transition. 
At the transition temperature $\tkt$, $m_3$ behaves 
as $m_3(\tkt)\sim L^{-0.128}$ (not shown). The exponent is very close to the KT value, $\beta/\nu=1/8$. 
In the bottom part of Fig.~\ref{fig.helicity}, the averaged data for the
helicity modulus $\Upsilon$~\cite{FiBaJa73} for $\Delta=0.25$ is shown along with the line $18T/\pi$.
Following Refs.~\cite{WeMi88,HuWe13}, 
we first fit the helicity modulus data for
a fixed temperature assuming a logarithmic approach to its asymptotic value:
\begin{equation}
\Upsilon_{\scriptstyle\rm fit}(L)=\frac{2TA}{\pi}\left( 1 + \frac{1}{2} \frac{1}{\log L + C}\right),
\end{equation}
where $A$ and $C$ are fitting parameters. The constant $A$ is related to the vorticity
of the system and is expected to be $A=1/\lambda^2$ at the transition. This form, valid
at $\tkt$, was inspired by renormalization group calculations and was observed to be valid 
even for very small systems~\cite{WeMi88}. Away from $\tkt$, the above expression is no
longer valid, the data deviates from it and the fitting error increases. Indeed, after repeating the process for
several temperatures close to the transition, the critical temperature corresponds to the one that 
minimizes the normalized quadratic error,
\begin{equation}
\varepsilon=\sum_i \left( \frac{\langle\Upsilon(T,L_i)\rangle-\Upsilon_{\scriptstyle\rm fit}(T,L_i)}{\sigma(T,L_i)}\right)^2,
\label{eq.erro}
\end{equation}
with $\sigma(T,L_i)=\sqrt{\langle\Upsilon^2\rangle -\langle\Upsilon\rangle^2}$. Following this
procedure we obtain, for $\Delta=0.25$, $T_{\scriptstyle\rm KT}\simeq 0.671$ and $A\simeq 8.97$ ($\lambda\simeq 1/3$). 
Within  error
bars, this value is the same as the lower bound.  For the F-P transition one expects $\lambda=1$
and, repeating the same procedure one gets, for $\Delta=0.7$, $A\simeq 0.99$ and $\tkt\simeq 0.748$.

\begin{figure}[htb]
\includegraphics[width=8cm]{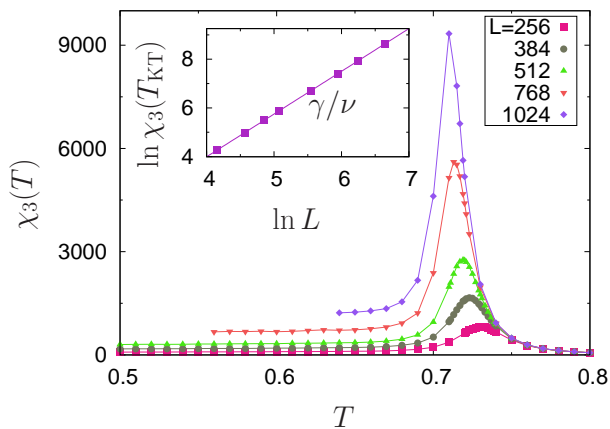}
\caption{The susceptibility $\chi_3(T)$ associated with $m_3$ near the KT transition for
$\Delta=0.25$.
Notice that $\chi_3$ grows with the system size even far below the critical temperature. At
the transition, the peak increases as $\chi_3(T_{\scriptstyle\rm KT})\sim L^{1.755}$,
where the exponent was obtained from the fit shown in the inset. }
\label{fig.susc}
\end{figure}

\begin{figure}[htb]
\includegraphics[width=8cm]{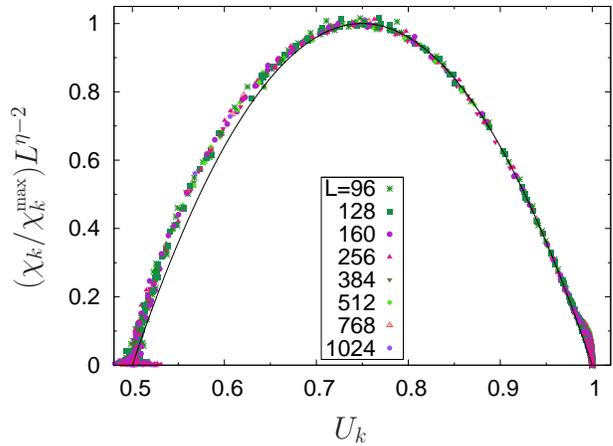}
\caption{Collapse of the rescaled susceptibility {\it vs.} the Binder cumulant, 
$U_k=\langle m_k^2\rangle^2/\langle m_k^4\rangle$, for several values of $\Delta$. With $k=3$
we have $\Delta=0.1$ and 0.25 and, with $k=1$, $\Delta=0.6$, 0.7, 0.9 and 1.
The linear size ranges from $L=96$ to 1024. The collapse is obtained with the KT value of
the exponent, $\eta=1/4$. In addition, by rescaling the curves with the maximum value of each
susceptibility, $\chi_k^{\scriptscriptstyle\rm max}(\Delta)$, the results for both P-N and P-F 
transitions all collapse on the same universal curve. 
The solid line, showing a small deviation from the parabolic behavior, is just a guide
to the eyes.}
\label{fig.binder}
\end{figure}

Fig.~\ref{fig.susc} shows that the susceptibility $\chi_3$ 
diverges for all temperatures below $\tkt$ as $L \rightarrow \infty$.
The exponent is clearly non universal and is larger in the
critical region. For a KT phase transition, for increasing $L$,
one expects $\chi_3(T_{\scriptstyle\rm KT})\sim L^{\gamma/\nu}$,
with $\gamma/\nu=7/4$. Indeed, we find $\chi_3(T_{\scriptstyle\rm KT})\sim L^{1.755}$,
as can be seen in the inset of Fig.~\ref{fig.susc}. Further evidence of the transition
can be obtained from the Binder cumulant, Eq.~(\ref{eq.binder}), even without the knowledge of 
the critical temperature. Near the phase transition the Binder cumulant scales as
$U=h(L/\xi)$, where $h(x)$ is a scaling function, and $\xi$ is the correlation length.  On the other hand $\chi_3=L^{2-\eta}g(L/\xi)$,
so that $\chi_3 L^{\eta-2}=g[h^{-1}(U)]$. Therefore, by 
plotting $\chi_3 L^{\eta-2}$ (with $\eta=1/4$ for KT transition) vs. the Binder cumulant~\cite{Loison99},
all the susceptibilities for different system sizes and temperatures should collapse onto one universal curve.
This is precisely what is found in our simulations.
Fig.~\ref{fig.binder} shows the data collapse for $U_1$ and $\chi_1$ for several values of $\Delta<0.5$ 
and $U_3$ and $\chi_3$ for $\Delta>0.5$. After rescaling the collapsed curves by the height of their maxima, 
$\chi_k^{\scriptscriptstyle\rm max}(\Delta)$, all points fall on the same universal curve close to the critical region (the line, a parabolic fit, is
just a guide to the eyes). A probable explanation is that, despite corresponding to transitions between different 
phases, P-N and P-F, they are all in the KT universality class.

\begin{figure}[htb]
\vskip 0.5\baselineskip
\includegraphics[width=8cm]{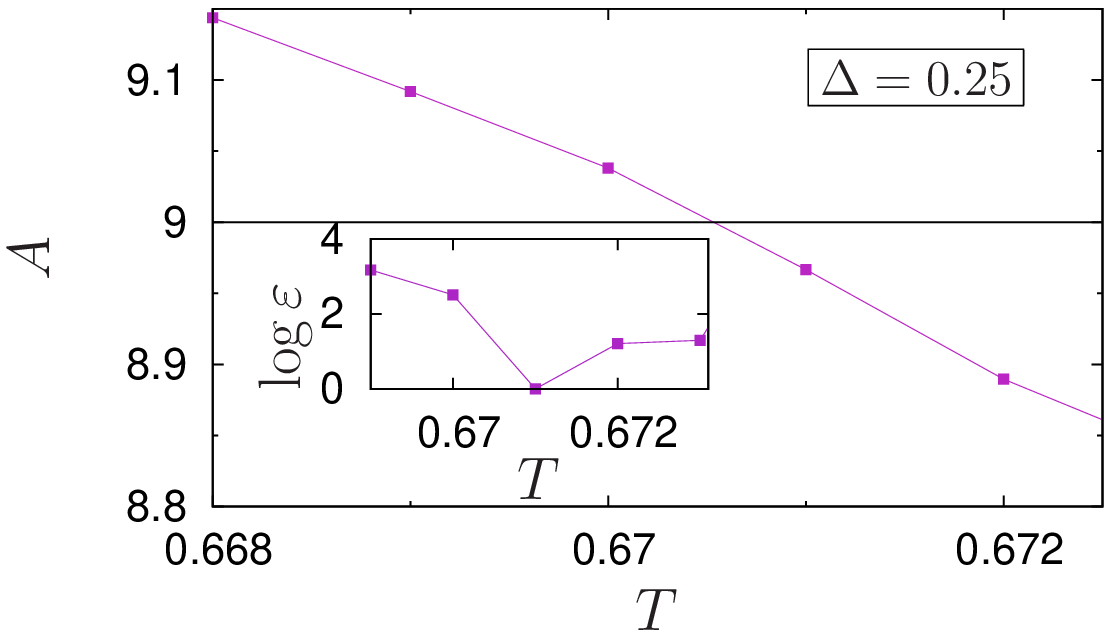}
\includegraphics[width=8cm]{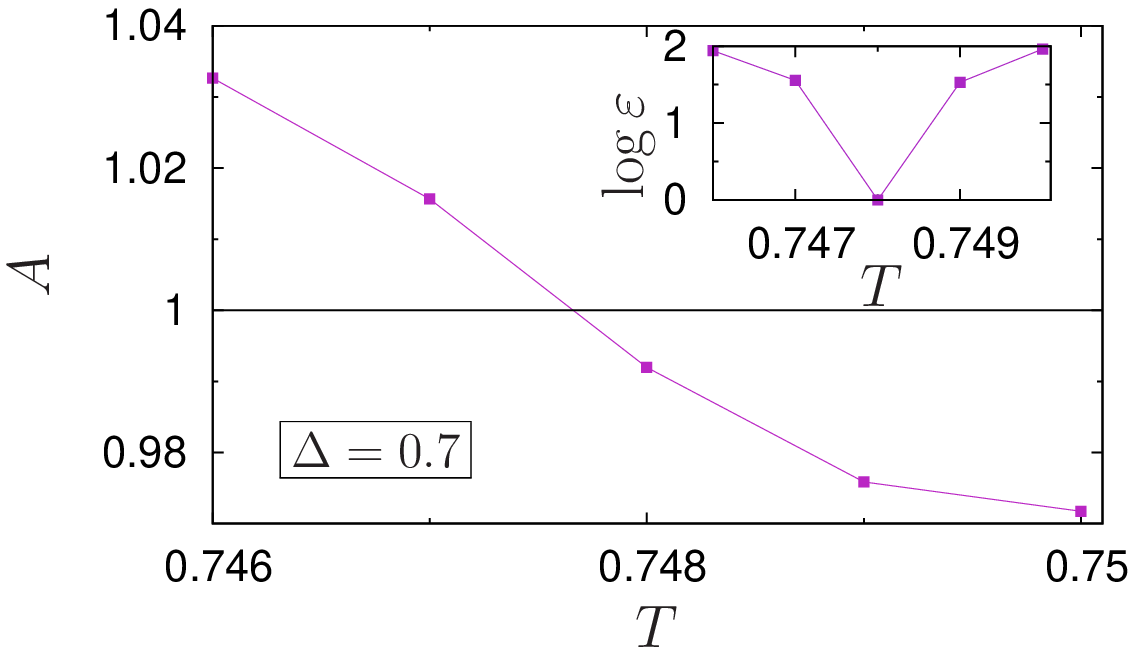}
\caption{Fit parameter $A$ of the helicity modulus for $\Delta=0.25$ (top) and
0.7 (bottom). Notice that $A$ crosses the
value 9 (top), that is, $q^2$, and 1 (bottom) very close to the temperature in which the fitting 
error is minimum.}
\label{fig.err}
\end{figure}

\begin{figure}[htb]
\includegraphics[width=8cm]{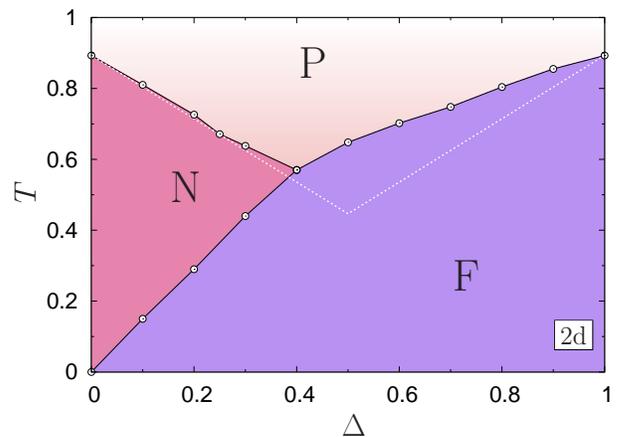}
\caption{Phase diagram for $q=3$. There are two KT phases, N and F,
both with quasi long-range order.  The points 
were obtained using the helicity modulus, while the lines are only guides
to the eye. The transition between N and F phases is in the 3-states Potts universality class.
The dashed lines are the lower bounds for the order-disorder transitions obtained in Ref. \cite{Romano06},
$T=\tkt(0)(1-\Delta)$ and $T=\tkt(0)\Delta$ for $\Delta\leq 0.5$ and
$\Delta\geq 0.5$, respectively, with a multicritical point at
$\dmult=1/2$ and $\tkt(1/2)=\tkt(0)/2$. }
\label{fig.diagrama}
\end{figure}

\begin{figure}[htb]
\includegraphics[width=8cm]{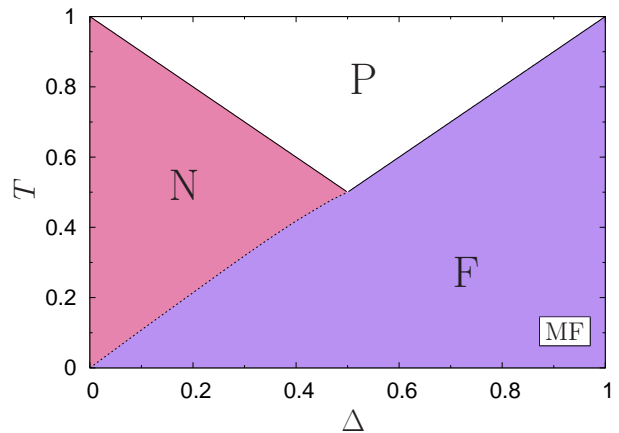}
\caption{Mean field phase diagram for $q=3$ generalized XY model. 
The P-N and P-F transitions phase are continuous (solid lines)
while the transition N-F (dashed line) is of first order.}
\label{fig.mfdiagrama}
\end{figure}

The information above can be used to construct the phase diagram for the $q=3$ generalized XY model shown in  
Fig.~\ref{fig.diagrama}.
Remarkably, the transition line N-P is very close to the lower bound
calculated in Ref.~\cite{Romano06} (except, close to the
multicritical point). The F-P line, on the other hand, is well above the lower bound
and, as a consequence, the multicritical point is located away from $\Delta=0.5$.
This phase diagram is qualitatively similar to the one obtained using 
a simple mean-field analysis of the Hamiltonian Eq.~(\ref{eq.H}). Within
the mean-field approximation all the spins are connected and the Mermin-Wagner theorem does not
apply.   The magnetizations, $m_1$ and $m_3$, become the true order parameters, and are found to 
satisfy a set of coupled equations 
\begin{align}
m_1 &= \frac{{\cal I}_1(2\beta m_1\Delta,2\beta m_3(1-\Delta))}{{\cal I}_0(2\beta m_1\Delta,2\beta m_3(1-\Delta))} \\
m_3 &= \frac{{\cal I}_3(2\beta m_1\Delta,2\beta m_3(1-\Delta))}{{\cal I}_0(2\beta m_1\Delta,2\beta m_3(1-\Delta))},
\label{ms}
\end{align}
where $\beta=1/T$, ${\cal I}_1(x,y)=\partial {\cal I}_0(x,y)/\partial x$, ${\cal I}_3(x,y)=\partial {\cal I}_0(x,y)/\partial y$
and
\begin{equation}
{\cal I}_0(x,y)=\frac{1}{2\pi}\int_0^{2\pi}\!\! d\theta  \exp\left[ x\cos\theta
                                                         +y\cos(3\theta)\right].
\end{equation}
The free energy density is
\begin{equation}
f = -T\ln {\cal I}_0(2\beta m_1\Delta,2\beta m_3(1-\Delta)) + m_1^2\Delta + (1-\Delta)m_3^2.
\end{equation}
At high temperatures there is a paramagnetic phase with $m_1=m_3=0$. The 
phase transition between the 
P and F phases occurs at
$T_c=\Delta$ for $\Delta\geq 0.5$  and the transition between the P and N phases happens at $T_c=1-\Delta$ for $\Delta\leq 0.5$, 
see Fig.~\ref{fig.mfdiagrama}.  Both transitions are of second order. Within the N phase, $m_1$ is identically zero, while the nematic order vanishes
as $m_3 \sim (T_c-T)^{1/2}$ as the phase transition line is approached from below.  
On the other hand, inside the F phase the order parameters vanish as
$m_1 \sim (T_c-T)^{1/2}$ and $m_3 \sim (T_c-T)^{3/2}$ as the F-P phase  boundary is approached from below.
Notice that the critical temperature for $\Delta=0$ (and 1) is $T=1$ and differ
from the (smaller) KT value. As expected the fluctuations decrease the critical temperature. 
At lower temperatures, and $0\leq \Delta\leq 0.5$, there is a second transition  
at which $m_1$ jumps discontinuously from 0 to a finite value, corresponding to a first order transition between
the N and F phases.  Thus, although the nature of the phase
transitions is not correctly captured by the mean field theory, both the topology
of the phase diagram and the fact that the transition N-F is not in the same universality class
as the  transitions between P-N and P-F phases is correctly predicted.

\section{Discussion and Conclusions}
\label{section.conclusions}

In conclusion, we have studied, through extensive Monte Carlo simulations 
and FSS, a generalized XY model with $q=3$. Contrary to the early
doubts regarding the existence of a nematic-like (N) phase in this model,
as opposed to a simple crossover~\cite{Korshunov12}, we
have presented strong numerical evidence that the model has P, N, and F phases. 
Curiously, the boundary between P and N phases coincides closely
with the lower bound obtained in Ref.~\cite{Romano06} while,
on the other hand, the phase transition between F and P phases
lies well above it.
The transition between the N and F phases, which for $q=2$ is in the
2d Ising universality class, for $q=3$ belongs to the 
3 states Potts universality class.  The overall topology of the $q=2$ and
3 cases are very similar and a simple mean-field analysis is
capable to grasp the existence (albeit not the actual nature) of
each phase.

With the possible exception of the region around the multicritical
point, the transition line N-P is very close to the lower bound
(indicated in Fig.~\ref{fig.mfdiagrama} by the dotted line)
predicted by Romano~\cite{Romano06}. That is, the declivity of $\tkt(\Delta)$,
within our precision, is $-1$. On the other hand, the line F-P has
a slope smaller than unity and is located far above the lower
bound. As a consequence, the multicritical point at which both
lines merge is not located at $\Delta=1/2$ and is above the lower
bound $\tkt/2$. These features can be observed in the phase
diagram Fig.~\ref{fig.diagrama} and in the diagram for $q=2$~\cite{HuWe13} 
as well. Moreover, for different values of $q$, the multicritical point 
$\dmult^{(q)}$ is located at increasing values of
$\Delta$: $\dmult^{(2)}\simeq 0.32$~\cite{HuWe13}, $\dmult^{(3)}\simeq 0.4$. 
For $q=8$, we do not yet have a precise location, but it appears to 
be close to $\Delta=0.5$.  We
observe that these points tend to the multicritical point consistent with the lower bound
calculated by Romano~\cite{Romano06}, $\dmult=1/2$. Thus, the F-P transition approaches,
as $q$ increases, the lower bound $T_c(\Delta)=T_{\scriptstyle\rm KT}\Delta $ (or, possibly,
there may exist a critical value of $q$ above which the lower bound might be exact).

There are several possible future extensions of this work. We are 
presently performing a detailed study of this model in 3d. 
In 2d it is important to explore the phase diagrams for larger 
$q$ values, since there are indications of the change in topology
that occurs for larger $q$. In particular, 
it was observed~\cite{PoArLe11} that, differently
from $q=2$ and 3, new phases appear for $q=8$.  A detailed
exploration of the nature of these phases is in order, not only
for $q=8$ but for intermediate values as well.
Finally, 
it has recently been proposed~\cite{ShLaFe11,BoWe12}, that for $q=2$, there may
exist a region close to the multicritical point in which the transition
to the paramagnetic phase is not KT, but  Ising. The very existence of
this Ising transition region is still an open question~\cite{HuWe13}
(and thus care must be taken with the use of the term {\it multicritical}). 
It will be interesting to see whether this topology might also extend to 
the larger $q$ values.

\begin{acknowledgments}
We thank S. Romano for bringing Ref.~\cite{Romano06} to our attention and F. C. Poderoso
for collaborating at the early stages of this project. JJA acknowledges the warm hospitality
of the LPTHE-Jussieu in Paris during his stay where part of this work was done and a
discussion with F. Zamponi on the mean field approach for this model.
This work was supported by CNPq, CAPES, FAPERGS, INCT-SC, INCT-FCx, 
and by US-AFOSR under the grant 
FA9550-12-1-0438.
\end{acknowledgments}

\bibliographystyle{apsrev}

\end{document}